\documentclass[a4paper,11pt]{article}
\pdfoutput=1
\usepackage{jheppub}

\preprint{YITP-21-81, TU-1132}

\title{\boldmath Semiclassical analysis of axion-assisted and axion-driven pair production}

\author[a]{Hiroyuki Kitamoto}
\author[b,c]{and Masaki Yamada}

\affiliation[a]{Yukawa Institute for Theoretical Physics, Kyoto University, 
Kyoto 606-8502, Japan}
\affiliation[b]{Department of Physics, Tohoku University,
Sendai, Miyagi 980-8578, Japan}
\affiliation[c]{Frontier Research Institute for Interdisciplinary Sciences, Tohoku University,
Sendai, Miyagi 980-8578, Japan}

\emailAdd{hiroyuki.kitamoto@yukawa.kyoto-u.ac.jp}
\emailAdd{m.yamada@tohoku.ac.jp}

\abstract{
We study the pair production of fermions in a time dependent axion background 
with and without an electric background. 
We construct the adiabatic mode functions 
which incorporate the gauge field and the axion velocity dependence of the dispersion relation. 
The semiclassical approach using this adiabatic basis shows two types of pair production. 
One is \textit{axion-assisted pair production}: 
the presence of the axion velocity gives enhancement and interference effects 
on the pair production driven by the electric field. 
The other is \textit{axion-driven pair production}: 
the time variation of the axion velocity causes the pair production 
even though the electric field is absent. 
}

\makeatletter
\gdef\@fpheader{}
\makeatother

\begin{document}

\maketitle

\flushbottom

\section{Introduction}\label{S-Intro}

The presence of a strong electric background causes 
the pair production of charged particles \cite{Schwinger:1951nm}. 
The so-called Schwinger effect is the most well known example of pair production. 
We here note that some cosmological scenarios treat electromagnetic fields 
with time evolution of an axion (or an axion-like particle), 
which is a pseudo-Nambu-Goldstone boson coupled to a photon via an anomaly. 
As an example, in the scenario of the magnetogenesis 
by the axion inflation \cite{Turner:1987bw, Garretson:1992vt, Anber:2006xt, Barnaby:2011qe, 
Domcke:2018eki, Kamada:2019uxp, Domcke:2019qmm}, 
a large electric field as well as a magnetic field is generated 
by an axion velocity via the Chern-Simons coupling. 

Motivated by the application to the above scenario, 
Domcke \textit{et al.} studied the pair production of charged fermions 
in a constant electric field and a constant axion velocity \cite{Domcke:2021fee}.\footnote{ 
They also studied the more appropriate situation for the magnetogenesis scenario, 
where a constant magnetic field is present. 
The presence of the magnetic field discretizes quantum numbers
while does not change the roles of the axion velocity in the axion-assisted Schwinger effect.
In this paper, for simplicity, we discuss the case where the magnetic field is absent.}
They found that the presence of the axion velocity
enhances the amplitude of the distribution function of produced fermions, 
and the distribution function shows an oscillatory behavior with the change of the axion velocity.
They named this phenomenon axion-assisted Schwinger effect, 
and empirically proposed an analytical formula 
which may describe the distribution function of produced fermions. 

In this paper, 
we derive the analytical formula for the axion-assisted Schwinger effect by using 
the semiclassical approach \cite{Brezin:1970xf, Popov:1972, Berry:1982, Kluger:1998bm, Dumlu:2011rr}. 
The fundamental strategy of the semiclassical approach is as follows. 
We expand the Dirac field by using the lowest-order WKB solutions 
(we call them the adiabatic mode functions in this paper) as a basis. 
The Bogoliubov coefficients are given by the coefficients of the adiabatic basis, and evolve with time 
as long as the backgrounds included explicitly in the adiabatic basis are time dependent. 

The semiclassical approach is used also in \cite{Domcke:2021fee}, 
while the adiabatic basis adopted in the previous study 
does not include the axion velocity dependence explicitly. 
In this paper, we construct the adiabatic mode functions 
which incorporate the gauge field and the axion velocity dependence of the dispersion relation.
Our semiclassical approach, using the axion velocity-modified adiabatic basis, 
clearly shows the roles of the axion velocity in the axion-assisted Schwinger effect, 
and gives the analytical formula for this effect. 

Furthermore, our semiclassical approach shows another type of pair production. 
If the axion velocity changes with time, the pair production occurs even though the electric field is absent. 
We here call this phenomenon axion-driven pair production. 
For the axion-driven pair production, the distribution function of produced fermions 
is consistent with that found in \cite{Adshead:2015kza, Hook:2019zxa}. 
This mechanism can be particularly important 
for the kinetic misalignment mechanism proposed in \cite{Co:2019wyp, Co:2019jts}, 
where large axion velocity and axion acceleration are induced in the early Universe. 

The organization of this paper is as follows. 
In Sec. \ref{S-SA}, we construct the adiabatic mode functions 
with the gauge field and the axion velocity dependence, 
and thus derive the equation for the Bogoliubov coefficients 
based on the axion velocity-modified adiabatic basis. 
In Sec. \ref{S-AAPP}, 
we consider the case where the electric field is present and the axion velocity is constant, 
and study the axion-assisted pair production (the axion-assisted Schwinger effect). 
In Sec. \ref{S-ADPP}, 
we consider the case where the electric field is absent and the axion velocity changes with time, 
and study the axion-driven pair production. 
Sec. \ref{S-Con} is devoted for conclusion and discussion. 

\section{Semiclassical approach}\label{S-SA}

We consider a Dirac field $\psi$ 
which couples to a $U(1)$ gauge field $A_\mu$ and an axion(-like particle) $\phi$ 
through the following nonlinear terms, 
\begin{align}
S_\text{D}=\int d^4x \big[i\bar{\psi}\gamma^\mu(\partial_\mu-ieA_\mu)\psi 
+\bar{\psi}\gamma^\mu\gamma^5\partial_\mu\theta_5\psi
-m\bar{\psi}e^{2i\theta_m\gamma^5}\psi\big], 
\label{action1}\end{align}
where $\theta_5=c_5\phi/f_a$ and $\theta_m=c_m\phi/f_a$, 
$f_a$ is the axion decay constant, and $c_5$, $c_m$ are the dimensionless coupling constants. 
See Appendix \ref{A-Notation} for our notation of the gamma matrices. 

We consider an expanding universe with a scale factor $a$, 
while the scale factor dependence has been almost eliminated in (\ref{action1}) 
by the conformal transformation $a^{3/2}\psi\to\psi$. 
Unless otherwise stated, 
the coordinates are given by the conformally flat ones (i.e., $x^0$ is given by the conformal time $\tau$), 
and the Lorentz indices are lowered and raised 
by $\eta_{\mu\nu}=\text{diag}(-1,1,1,1)$ and its inverse $\eta^{\mu\nu}$, respectively. 
Although the mass term is time dependent as $m=am_\text{phys}$ 
($m_\text{phys}$ is the physical scale of the mass), 
we assume that the pair production occurs at the short time scale 
where the time evolution of the mass term is negligible.\footnote{
Specifically, we consider the parameter region 
$-e\partial_0A\ \text{or}\ \partial_0^2\theta \gg m\partial_0a/a$. 
Here, we compare the time derivatives of the backgrounds 
which appear linearly in the Dirac equation (\ref{Dirac}).} 

Using the chiral transformation $\psi\to e^{-i\theta_m\gamma^5}\psi$, 
the two axion couplings in (\ref{action1}) are unified as follows 
\begin{align}
S_\text{D}=\int d^4x \big[i\bar{\psi}\gamma^\mu(\partial_\mu-ieA_\mu)\psi 
+\bar{\psi}\gamma^\mu\gamma^5\partial_\mu\theta\psi
-m\bar{\psi}\psi\big], 
\label{action2}\end{align}
where $\theta=\theta_5+\theta_m$. 
In the subsequent discussion, $\psi$ denotes this conformal and chiral transformed Dirac field. 
In this paper, the gauge field and the axion are treated at the background level. 
Specifically, we consider the situation where these background fields are homogeneous, 
\begin{align}
\theta=\theta(\tau), \quad A_\mu=\delta_\mu^{\ 3}A(\tau). 
\end{align}
The action (\ref{action2}) includes the axion velocity $\partial_0\theta$ rather than $\theta$ itself. 
We assume that the axion velocity does not change its sign 
in the time range of our interest. 
In such a situation, we can set it to be positive, $\partial_0\theta\ge 0$ without loss of generality. 

After the spatial Fourier transformation, the Dirac equation is given by 
\begin{align}
(i\gamma^0\partial_0-\gamma^i \bar{k}_i+\gamma^0\gamma^5\partial_0\theta-m)\psi_\mathbf{k}(\tau)=0, 
\label{Dirac}\end{align}
where $\bar{k}_i=(k_1,k_2,k_3-eA)$ and $k_i$ is the comoving momentum. 
Inserting the following equation 
\begin{align}
\psi_\mathbf{k}(\tau)
=(i\gamma^0\partial_0-\gamma^i \bar{k}_i-\gamma^0\gamma^5\partial_0\theta+m)\hat{\psi}_\mathbf{k}(\tau), 
\label{h-psi}\end{align}
we obtain the Klein-Gordon like equation 
\begin{align}
\big[-\partial_0^2-\bar{k}^2-m^2 
+(\partial_0\theta)^2-2i\gamma^5\partial_0\theta\partial_0&+2m\gamma^0\gamma^5\partial_0\theta \notag\\
&+ie\gamma^0\gamma^3 F_{03}-i\gamma^5\partial_0^2\theta\big]\hat{\psi}_\mathbf{k}=0, 
\label{KG}\end{align}
where $\bar{k}=\sqrt{\bar{k}_i\bar{k}^i}$ and $F_{\mu\nu}=\partial_\mu A_\nu-\partial_\nu A_\mu$. 

First, we derive the lowest-order WKB solutions of the Dirac equation 
(we call them the adiabatic mode functions in this paper). 
We can derive the adiabatic mode functions 
by treating $A_\mu$ and $\partial_0\theta$ as if they were constant. 
The Dirac equation (\ref{Dirac}) includes $A_\mu$ and $\partial_0\theta$, 
but does not include their time derivatives. 
For the Klein-Gordon like equation (\ref{KG}), 
the last two terms in the left-hand side are neglected in deriving the adiabatic mode functions. 

For the positive-frequency adiabatic mode functions, we replace the time derivative with the frequency, 
\begin{align}
\partial_0\to -i\omega_{\mathbf{k},\pm}. 
\label{replacement}\end{align}
Note that we have not determined the explicit forms of the frequencies. 
We introduced the labels $\pm$ to distinguish among the eigenvalues of the matrix term in (\ref{KG}), 
\begin{align}
2\partial_0\theta(-\gamma^5\omega_{\mathbf{k},\pm}+m\gamma^0\gamma^5)
\hat{\mu}_{\mathbf{k},\pm} 
=\pm 2\partial_0\theta\sqrt{\omega_{\mathbf{k},\pm}^2-m^2}\hat{\mu}_{\mathbf{k},\pm}, 
\label{mu1}\end{align}
\begin{align}
\hat{\mu}_{\mathbf{k},+}
&=\begin{pmatrix} \big(\omega_{\mathbf{k},+}+\sqrt{\omega_{\mathbf{k},+}^2-m^2}\big)\xi \\ 
-m\xi \end{pmatrix}, \notag\\
\hat{\mu}_{\mathbf{k},-}
&=\begin{pmatrix} -m\xi \\ 
\big(\omega_{\mathbf{k},-}+\sqrt{\omega_{\mathbf{k},-}^2-m^2}\big)\xi \end{pmatrix}, 
\label{mu2}\end{align}
where $\xi=(1,0)^T$. 
In fact, we do not need to consider all choices of two-spinors. 
In deriving the adiabatic mode functions, we multiply $\hat{\mu}_{\mathbf{k},\pm}$ 
by $(\gamma^0\omega_{\mathbf{k},\pm}-\gamma^i\bar{k}_i-\gamma^0\gamma^5\partial_0\theta+m)$ 
in (\ref{h-psi}), which have zero eigenvalues.   
The adiabatic mode functions consisted of $\eta=(0,1)^T$ are linearly dependent 
of those consisted of $\xi=(1,0)^T$. 

From (\ref{replacement}) and (\ref{mu1}), the Klein-Gordon like equation (\ref{KG}) reduces to 
\begin{align}
\omega_{\mathbf{k},\pm}^2-\bar{k}^2-m^2 
+(\partial_0\theta)^2\pm 2\partial_0\theta\sqrt{\omega_{\mathbf{k},\pm}^2-m^2}=0. 
\label{KG'}\end{align}
We thus obtain the explicit forms of the frequencies, 
\begin{align}
\omega_{\mathbf{k},\pm}(\tau)=\sqrt{(\bar{k}\mp\partial_0\theta)^2+m^2}, 
\label{DR}\end{align}
where we consider the parameter region $\bar{k}\ge \partial_0\theta$. 
In this parameter region, the positivity of $\sqrt{\omega_{\mathbf{k},\pm}^2-m^2}$ 
allows one solution for each sign choice in (\ref{KG'}).\footnote{
In the other parameter region $\bar{k}\le \partial_0\theta$, 
the positivity of $\sqrt{\omega_{\mathbf{k},\pm}^2-m^2}$ allows two solutions for the minus sign 
while does not allow any solution for the plus sign. 
See Appendix \ref{A-Universality} for more details.}  
The dispersion relation with the axion velocity dependence can be seen 
also in \cite{Domcke:2021fee, Adshead:2015kza, Hook:2019zxa}. 
The presence of the axion velocity resolves the degeneracy, 
\begin{align}
\omega_{\mathbf{k},+}<\omega_{\mathbf{k},-}. 
\label{RD}\end{align}
We thus call $\omega_{\mathbf{k},+}$ the lower frequency, 
and call $\omega_{\mathbf{k},-}$ the higher frequency. 
Recall that we set $\partial_0\theta$ to be positive. 
The sign flip of $\partial_0\theta$ just exchanges the roles of $\omega_{\mathbf{k},\pm}$. 

The positive-frequency adiabatic mode functions $u_{\mathbf{k},\pm}$ are given by 
\begin{align}
u_{\mathbf{k},+}
&\propto e^{-i\Theta_{\mathbf{k},+}}
(\gamma^0\omega_{\mathbf{k},+}-\gamma^i\bar{k}_i-\gamma^0\gamma^5\partial_0\theta+m)
\hat{\mu}_{\mathbf{k},+} \notag\\
&=\frac{e^{-i\Theta_{\mathbf{k},+}}}{N_{\mathbf{k},+}}
\begin{pmatrix} m(\bar{k}+\bar{k}_3)e^{-i\varphi_\perp/2} \\
mk_\perp e^{i\varphi_\perp/2} \\ 
(\omega_{\mathbf{k},+}+\bar{k}-\partial_0\theta)(\bar{k}+\bar{k}_3)e^{-i\varphi_\perp/2} \\ 
(\omega_{\mathbf{k},+}+\bar{k}-\partial_0\theta)k_\perp e^{i\varphi_\perp/2} \end{pmatrix}, 
\label{u+}\end{align}
\begin{align}
u_{\mathbf{k},-}
&\propto e^{-i\Theta_{\mathbf{k},-}}
(\gamma^0\omega_{\mathbf{k},-}-\gamma^i\bar{k}_i-\gamma^0\gamma^5\partial_0\theta+m)
\hat{\mu}_{\mathbf{k},-} \notag\\
&=\frac{e^{-i\Theta_{\mathbf{k},-}}}{N_{\mathbf{k},-}}
\begin{pmatrix} (\omega_{\mathbf{k},-}+\bar{k}+\partial_0\theta)k_\perp e^{-i\varphi_\perp/2} \\ 
-(\omega_{\mathbf{k},-}+\bar{k}+\partial_0\theta)(\bar{k}+\bar{k}_3)e^{i\varphi_\perp/2} \\ 
mk_\perp e^{-i\varphi_\perp/2} \\
-m(\bar{k}+\bar{k}_3)e^{i\varphi_\perp/2} \end{pmatrix}, 
\label{u-}\end{align}
where $k_\perp=\sqrt{k_1^2+k_2^2}$, $e^{i\varphi_\perp}=(k_1+ik_2)/k_\perp$, 
and the time dependent phase factors and the normalization factors are given by 
\begin{align}
\Theta_{\mathbf{k},\pm}=\int d\tau \omega_{\mathbf{k},\pm}, 
\end{align}
\begin{align}
N_{\mathbf{k},\pm}
=2\sqrt{\omega_{\mathbf{k},\pm}(\omega_{\mathbf{k},\pm}+\bar{k}\mp\partial_0\theta)
\bar{k}(\bar{k}+\bar{k}_3)}. 
\end{align}
For convenience, we introduced the constant phase factor $e^{-i\varphi_\perp/2}$ 
in the second lines of (\ref{u+}) and (\ref{u-}). 

In a similar way, where we replace $\partial_0$ with $i\omega_{\mathbf{k},\pm}$, 
the negative-frequency adiabatic mode functions $v_{\mathbf{k},\pm}$ are constructed as follows 
\begin{align}
v_{\mathbf{k},+}
&\propto e^{i\Theta_{\mathbf{k},+}}
(-\gamma^0\omega_{\mathbf{k},+}-\gamma^i\bar{k}_i-\gamma^0\gamma^5\partial_0\theta+m)
\hat{\nu}_{\mathbf{k},+} \notag\\
&=\frac{e^{i\Theta_{\mathbf{k},+}}}{\tilde{N}_{\mathbf{k},+}}
\begin{pmatrix} -(\omega_{\mathbf{k},+}+\bar{k}-\partial_0\theta)k_\perp e^{-i\varphi_\perp/2} \\
-(\omega_{\mathbf{k},+}+\bar{k}-\partial_0\theta)(\bar{k}-\bar{k}_3)e^{i\varphi_\perp/2} \\ 
mk_\perp e^{-i\varphi_\perp/2} \\
m(\bar{k}-\bar{k}_3)e^{i\varphi_\perp/2} \end{pmatrix}, 
\label{v+}\end{align}
\begin{align}
v_{\mathbf{k},-}
&\propto e^{i\Theta_{\mathbf{k},+}}
(-\gamma^0\omega_{\mathbf{k},+}-\gamma^i\bar{k}_i-\gamma^0\gamma^5\partial_0\theta+m)
\hat{\nu}_{\mathbf{k},-} \notag\\
&=\frac{e^{i\Theta_{\mathbf{k},-}}}{\tilde{N}_{\mathbf{k},-}}
\begin{pmatrix} m(\bar{k}-\bar{k}_3)e^{-i\varphi_\perp/2} \\
-mk_\perp e^{i\varphi_\perp/2} \\ 
-(\omega_{\mathbf{k},-}+\bar{k}+\partial_0\theta)(\bar{k}-\bar{k}_3)e^{-i\varphi_\perp/2} \\
(\omega_{\mathbf{k},-}+\bar{k}+\partial_0\theta)k_\perp e^{i\varphi_\perp/2} \end{pmatrix}, 
\label{v-}\end{align}
where $\hat{\nu}_{\mathbf{k},\pm}$ are the eigenvectors of the matrix term in (\ref{KG}), 
\begin{align}
2\partial_0\theta(\gamma^5\omega_{\mathbf{k},\pm}+m\gamma^0\gamma^5)
\hat{\nu}_{\mathbf{k},\pm} 
=\pm 2\partial_0\theta\sqrt{\omega_{\mathbf{k},\pm}^2-m^2}\hat{\nu}_{\mathbf{k},\pm}, 
\label{nu1}\end{align}
\begin{align}
\hat{\nu}_{\mathbf{k},+}
&=\begin{pmatrix} m\xi \\ 
\big(\omega_{\mathbf{k},+}+\sqrt{\omega_{\mathbf{k},+}^2-m^2}\big)\xi \end{pmatrix}, \notag\\
\hat{\nu}_{\mathbf{k},-}
&=\begin{pmatrix} \big(\omega_{\mathbf{k},-}+\sqrt{\omega_{\mathbf{k},-}^2-m^2}\big)\xi \\
m\xi \end{pmatrix}, 
\label{nu2}\end{align}
and the normalization factors are given by 
\begin{align}
\tilde{N}_{\mathbf{k},\pm}
&=2\sqrt{\omega_{\mathbf{k},\pm}(\omega_{\mathbf{k},\pm}+\bar{k}\mp\partial_0\theta)
\bar{k}(\bar{k}-\bar{k}_3)} \notag\\
&=N_{-\mathbf{k},\pm}|_{e\to-e}. 
\end{align}
For convenience, we introduced the constant phase factor $e^{-i\varphi_\perp/2}$ 
in the second lines of (\ref{v+}) and (\ref{v-}). 
Although we derived (\ref{DR}), (\ref{u+}), (\ref{u-}), (\ref{v+}), and (\ref{v-}) for $\bar{k}\ge \partial_0\theta$, 
the same frequencies and the same adiabatic mode functions hold true 
regardless of the sign of $(\bar{k}-\partial_0\theta)$. 
See Appendix \ref{A-Universality} for this universality. 

These adiabatic mode functions satisfy the following equations by definition, 
\begin{align}
&(\gamma^0\omega_{\mathbf{k},s}-\gamma^i\bar{k}_i+\gamma^0\gamma^5\partial_0\theta-m)
u_{\mathbf{k},s}=0, \notag\\
&(-\gamma^0\omega_{\mathbf{k},s}-\gamma^i\bar{k}_i+\gamma^0\gamma^5\partial_0\theta-m)
v_{\mathbf{k},s}=0, 
\label{Adiabatic}\end{align}
and they are orthonormal, 
\begin{align}
u_{\mathbf{k},s}^\dagger u_{\mathbf{k},s'}=\delta_{ss'}, \quad
v_{\mathbf{k},s}^\dagger v_{\mathbf{k},s'}=\delta_{ss'}, \quad
v_{\mathbf{k},s}^\dagger u_{\mathbf{k},s'}=0. 
\label{orthonormal}\end{align}
The positive and the negative frequency modes are exchanged 
by the charge conjugate transformation $\chi^C\equiv\gamma^2\chi^*$ 
(where $\chi$ is an arbitrary four-spinor), 
\begin{align}
u_{-\mathbf{k},s}^C|_{e\to -e}=v_{\mathbf{k},s}, \quad
v_{-\mathbf{k},s}^C|_{e\to -e}=u_{\mathbf{k},s}, 
\end{align}
where we define that $\varphi_\perp$ in $u_{\mathbf{k},s}$ shifts by $+\pi$ 
while $\varphi_\perp$ in $v_{\mathbf{k},s}$ shifts by $-\pi$ under $\mathbf{k}\to -\mathbf{k}$. 
In order to obtain this simple relation, 
we introduced $e^{-i\varphi_\perp/2}$ into the adiabatic mode functions. 

Second, as with \cite{Brezin:1970xf, Popov:1972, Berry:1982, Kluger:1998bm, Dumlu:2011rr}, 
we expand the Dirac field by using the adiabatic mode functions as a basis. 
At the initial time $\tau_\text{ini}$, we introduce the annihilation and the creation operators as follows 
\begin{align}
\psi(\tau_\text{ini},\mathbf{x})
=\int\frac{d^3k}{(2\pi)^3}\sum_{\sigma=\pm}
\big[a_{\mathbf{k},\sigma}u_{\mathbf{k},\sigma}(\tau_\text{ini}) 
+b_{-\mathbf{k},\sigma}^\dagger v_{\mathbf{k},\sigma}(\tau_\text{ini})\big]
e^{i\mathbf{k}\cdot\mathbf{x}}, 
\label{initial1}\end{align}
\begin{align}
&\{a_{\mathbf{k},\sigma}, a_{\mathbf{k}',\sigma'}^\dagger\} 
=\{b_{\mathbf{k},\sigma}, b_{\mathbf{k}',\sigma'}^\dagger\} 
=(2\pi)^3\delta(\mathbf{k}-\mathbf{k}')\delta_{\sigma\sigma'}, \notag\\
&\{a_{\mathbf{k},\sigma}, a_{\mathbf{k}',\sigma'}\} 
=\{b_{\mathbf{k},\sigma}, b_{\mathbf{k}',\sigma'}\} 
=\{a_{\mathbf{k},\sigma}, b_{\mathbf{k}',\sigma'}\} 
=\{a_{\mathbf{k},\sigma}, b_{\mathbf{k}',\sigma'}^\dagger\}=0, 
\end{align}
where we use $\sigma$ rather than $s$ 
to emphasize that this label is defined independently of time. 
The Dirac field at an arbitrary time can be expressed as follows 
\begin{align}
\psi(x)=\int\frac{d^3k}{(2\pi)^3}\sum_{\sigma=\pm}
\big[a_{\mathbf{k},\sigma}\psi_{\mathbf{k},\sigma}(\tau) 
+b_{-\mathbf{k},\sigma}^\dagger\psi_{-\mathbf{k},\sigma}^C(\tau)|_{e\to -e}\big]
e^{i\mathbf{k}\cdot\mathbf{x}}, 
\end{align}
\begin{align}
\psi_{\mathbf{k},\sigma}(\tau)
=\sum_{s=\pm}\big[\alpha_{\mathbf{k},s\sigma}(\tau)u_{\mathbf{k},s}(\tau) 
+\beta_{\mathbf{k},s\sigma}(\tau)v_{\mathbf{k},s}(\tau)\big], 
\label{Bogoliubov1}\end{align}
where the Bogoliubov coefficients $\alpha_{\mathbf{k},s\sigma}$, $\beta_{\mathbf{k},s\sigma}$ 
take the following initial vales 
\begin{align}
\alpha_{\mathbf{k},s\sigma}(\tau_\text{ini})=\delta_{s\sigma}, \quad
\beta_{\mathbf{k},s\sigma}(\tau_\text{ini})=0. 
\label{initial2}\end{align}
Unless both $A_\mu$ and $\partial_0\theta$ are constant, 
the adiabatic mode functions do not satisfy the Dirac equation (\ref{Dirac}), 
and instead satisfy (\ref{Adiabatic}). 
Since the left-hand side of (\ref{Bogoliubov1}) satisfies the Dirac equation, 
the Bogoliubov coefficients evolve with time 
such that the right-hand side satisfies the Dirac equation. 
In other words, the pair production occurs 
as long as the electric field or the axion acceleration is nonzero, 
\begin{align}
E\equiv-\partial_0A\not=0 \quad
\text{or} \quad \partial_0^2\theta\not=0. 
\end{align}

Let us derive the equation for the Bogoliubov coefficients. 
From (\ref{orthonormal}) and (\ref{Bogoliubov1}), the Bogoliubov coefficients are given by 
\begin{align}
\alpha_{\mathbf{k},s\sigma}=u_{\mathbf{k},s}^\dagger \psi_{\mathbf{k},\sigma}, \quad
\beta_{\mathbf{k},s\sigma}=v_{\mathbf{k},s}^\dagger \psi_{\mathbf{k},\sigma}. 
\label{Bogoliubov2}\end{align}
Furthermore, using (\ref{Dirac}) and (\ref{Adiabatic}), 
their time derivatives are expressed as follows 
\begin{align}
\partial_0\alpha_{\mathbf{k},s\sigma}
&=-i\omega_{\mathbf{k},s}\alpha_{\mathbf{k},s\sigma} 
+\sum_{s'=\pm}(\partial_0u_{\mathbf{k},s}^\dagger u_{\mathbf{k},s'}\alpha_{\mathbf{k},s'\sigma}
+\partial_0u_{\mathbf{k},s}^\dagger v_{\mathbf{k},s'}\beta_{\mathbf{k},s'\sigma}), \notag\\
\partial_0\beta_{\mathbf{k},s\sigma}
&=i\omega_{\mathbf{k},s}\beta_{\mathbf{k},s\sigma} 
+\sum_{s'=\pm}(\partial_0v_{\mathbf{k},s}^\dagger u_{\mathbf{k},s'}\alpha_{\mathbf{k},s'\sigma}
+\partial_0v_{\mathbf{k},s}^\dagger v_{\mathbf{k},s'}\beta_{\mathbf{k},s'\sigma}). 
\label{Bogoliubov3}\end{align}
We list the explicit forms of the products of the adiabatic mode functions and their time derivatives 
in Appendix \ref{A-Products}. 
Substituting (\ref{Products}) into (\ref{Bogoliubov3}), we obtain 
\begin{align}
\partial_0
\begin{pmatrix}
\alpha_{\mathbf{k},+\sigma} \\
\beta_{\mathbf{k},+\sigma} \\
\alpha_{\mathbf{k},-\sigma} \\
\beta_{\mathbf{k},-\sigma} 
\end{pmatrix}
=
&\left[eE
\begin{pmatrix}
0 &
\frac{m\bar{k}_3}{2\omega_{\mathbf{k},+}^2\bar{k}}e^{2i\Theta_{\mathbf{k},+}} &
C_\mathbf{k}e^{i(\Theta_{\mathbf{k},+}-\Theta_{\mathbf{k},-})} &
-D_\mathbf{k}e^{i(\Theta_{\mathbf{k},+}+\Theta_{\mathbf{k},-})} \\
-\frac{m\bar{k}_3}{2\omega_{\mathbf{k},+}^2\bar{k}}e^{-2i\Theta_{\mathbf{k},+}} &
0 &
-D_\mathbf{k}e^{-i(\Theta_{\mathbf{k},+}+\Theta_{\mathbf{k},-})} &
-C_\mathbf{k}e^{-i(\Theta_{\mathbf{k},+}-\Theta_{\mathbf{k},-})} \\
-C_\mathbf{k}e^{-i(\Theta_{\mathbf{k},+}-\Theta_{\mathbf{k},-})} &
D_\mathbf{k}e^{i(\Theta_{\mathbf{k},+}+\Theta_{\mathbf{k},-})} &
0 &
\frac{m\bar{k}_3}{2\omega_{\mathbf{k},-}^2\bar{k}}e^{2i\Theta_{\mathbf{k},-}} \\
D_\mathbf{k}e^{-i(\Theta_{\mathbf{k},+}+\Theta_{\mathbf{k},-})} &
C_\mathbf{k}e^{i(\Theta_{\mathbf{k},+}-\Theta_{\mathbf{k},-})} &
-\frac{m\bar{k}_3}{2\omega_{\mathbf{k},-}^2\bar{k}}e^{-2i\Theta_{\mathbf{k},-}} &
0  
\end{pmatrix}\right. \notag\\
&\left.+\partial_0^2\theta
\begin{pmatrix}
0 & -\frac{m}{2\omega_{\mathbf{k},+}^2}e^{2i\Theta_{\mathbf{k},+}} & 0 & 0 \\
\frac{m}{2\omega_{\mathbf{k},+}^2}e^{-2i\Theta_{\mathbf{k},+}} & 0 & 0 & 0 \\
0 & 0 & 0 & \frac{m}{2\omega_{\mathbf{k},-}^2}e^{2i\Theta_{\mathbf{k},-}} \\
0 & 0 & -\frac{m}{2\omega_{\mathbf{k},-}^2}e^{-2i\Theta_{\mathbf{k},-}} & 0 
\end{pmatrix}\right] 
\begin{pmatrix}
\alpha_{\mathbf{k},+\sigma} \\
\beta_{\mathbf{k},+\sigma} \\
\alpha_{\mathbf{k},-\sigma} \\
\beta_{\mathbf{k},-\sigma} 
\end{pmatrix}. 
\label{Bogoliubov4}\end{align}
See (\ref{C}) and (\ref{D}) for the explicit forms of $C_\mathbf{k}$ and $D_\mathbf{k}$. 
Note that the diagonal elements in the right-hand side are zero. 
The absence properly reflects that the pair production means the mixing of different modes with time, 
rather than the time evolution of each mode.

From (\ref{Bogoliubov4}) and (\ref{initial2}), 
we can confirm that the following unitary condition holds true, 
\begin{align}
\sum_{s=\pm}(|\alpha_{\mathbf{k},s\sigma}|^2+|\beta_{\mathbf{k},s\sigma}|^2)=1. 
\end{align}
In general, each $(|\alpha_{\mathbf{k},s\sigma}|^2+|\beta_{\mathbf{k},s\sigma}|^2)$ is not conserved, 
and $|\beta_{\mathbf{k},s\sigma}|^2$ is not equal to the distribution function of produced fermions. 
This is because the equation (\ref{Bogoliubov4}) 
has the block-off-diagonal elements between the lower and the higher frequency modes. 
However, as discussed in the next section, 
we may practically neglect the contribution from the higher-frequency modes 
including the block-off-diagonal elements. 

The equation (\ref{Bogoliubov4}) shows two types of pair production.  
If the axion velocity is constant, 
the pair production is driven by the electric field. 
The presence of the axion velocity modifies the electric pair production 
because the matrix with the coefficient $eE$ depends on $\partial_0\theta$ 
(as discussed in the next section, it enhances the amplitude of the distribution function). 
If the axion velocity changes with time, the pair production occurs 
even though the electric field is absent; i.e., it is driven by the axion acceleration. 
We call the former one \textit{axion-assisted pair production} 
(we also call it \textit{axion-assisted Schwinger effect} by following \cite{Domcke:2021fee}), 
and call the latter one \textit{axion-driven pair production}. 
In the subsequent sections, we study these two types of pair production more specifically. 

\section{Axion-assisted pair production}\label{S-AAPP}

We consider the case where the electric field is nonzero 
while the axion acceleration is zero in the whole time range, 
\begin{align}
E\not=0, \quad \partial_0^2\theta=0. 
\label{case1}\end{align}
The electric field is set to be constant and positive for simplicity. 
We consider a sizeable axion velocity such that the axion-assisted effect becomes manifest, 
\begin{align}
\partial_0\theta\gg k_\perp,\ m. 
\label{sizeable}\end{align}
Note, however, that we impose the additional condition (\ref{VC1}), 
which ensures the frequency difference between the adiabatic modes. 
The additional condition does not allow the trivial limit of (\ref{sizeable}): $k_\perp=m=0$. 

In this case, the pair production is driven by the first term in the right-hand side of (\ref{Bogoliubov4}). 
It is difficult to solve the differential equation with the $4\times 4$ matrix exactly. 
However, the presence of the axion velocity resolves the degeneracy as seen in (\ref{RD}). 
Since the contribution from the lower-frequency modes may be dominant 
compared with that from the higher-frequency modes, 
we focus on the former one: 
\begin{align}
\alpha_{\mathbf{k},s\sigma}
\simeq \alpha_{\mathbf{k},+}\delta_{s+}\delta_{\sigma+}+\delta_{s-}\delta_{\sigma-}, \quad
\beta_{\mathbf{k},s\sigma}
\simeq \beta_{\mathbf{k},+}\delta_{s+}\delta_{\sigma+}. 
\label{reduction}\end{align}
In this approximation, the unitary condition holds true for the lower-frequency modes, 
\begin{align}
|\alpha_{\mathbf{k},+}|^2+|\beta_{\mathbf{k},+}|^2\simeq 1, 
\end{align}
and the distribution function for the lower-frequency modes is given by 
\begin{align}
n_{\mathbf{k},+}\simeq |\beta_{\mathbf{k},+}|^2. 
\label{reduced-n}\end{align} 
Extracting the upper left block from the $4\times 4$ matrix 
(i.e., neglecting the block-off-diagonal elements), 
we obtain the reduced equation for $(\alpha_{\mathbf{k},+},\beta_{\mathbf{k},+})$, 
\begin{align}
\partial_0\begin{pmatrix} \alpha_{\mathbf{k},+} \\ \beta_{\mathbf{k},+} \end{pmatrix} 
\simeq 
eE\begin{pmatrix} 0 & \frac{m\bar{k}_3}{2\omega_{\mathbf{k},+}^2\bar{k}}e^{2i\Theta_{\mathbf{k},+}} \\
-\frac{m\bar{k}_3}{2\omega_{\mathbf{k},+}^2\bar{k}}e^{-2i\Theta_{\mathbf{k},+}} & 0 \end{pmatrix} 
\begin{pmatrix} \alpha_{\mathbf{k},+} \\ \beta_{\mathbf{k},+} \end{pmatrix}. 
\label{AAPP1}\end{align}
In the subsequent paragraphs, we discuss in which parameter region the reduction process is justified, 
and how to obtain the analytical solution of (\ref{AAPP1}). 

In this paper, we consider the situation where the semiclassical picture is valid on the real time axis, 
\begin{align}
\big(\frac{X_\mathbf{k}}{\omega_{\mathbf{k},+}}\big)^2\ll 1, \quad
\big|\frac{\partial_0 X_\mathbf{k}}{\omega_{\mathbf{k},+}^2}\big|\ll 1, 
\label{semiclassical}\end{align}
where $X_\mathbf{k}$ are the matrix elements without the oscillating parts 
(we consider the matrix elements which are relevant with the lower-frequency modes). 
The condition (\ref{semiclassical}) can be understood as the adiabaticity condition, 
which ensures the instantaneous particle picture at a given time.\footnote{
In short, (\ref{semiclassical}) means that the time evolution of the Bogoliubov coefficients is slow
compared with that of the adiabatic basis.
For scalar fields, the equation for the Bogoliubov coefficients is given by 
$\partial_0\begin{pmatrix} \alpha_\mathbf{k} \\ \beta_\mathbf{k} \end{pmatrix}
=\begin{pmatrix} 0 & \frac{\partial_0\omega_\mathbf{k}}{2\omega_\mathbf{k}}e^{2i\Theta_\mathbf{k}} \\
\frac{\partial_0\omega_\mathbf{k}}{2\omega_\mathbf{k}}e^{-2i\Theta_\mathbf{k}} & 0 \end{pmatrix}
\begin{pmatrix} \alpha_\mathbf{k} \\ \beta_\mathbf{k} \end{pmatrix}$ (see, e.g., \cite{Kluger:1998bm}), 
and thus the validity condition (\ref{semiclassical}) reduces to the well known one: 
$\big(\frac{\partial_0\omega_\mathbf{k}}{\omega_\mathbf{k}^2}\big)^2\ll 1$, 
$\big|\frac{\partial_0^2\omega_\mathbf{k}}{\omega_\mathbf{k}^3}\big|\ll 1$.} 
As we see in the subsequent discussion, 
it leads to the pair production whose distribution function is exponentially small. 
In such a situation, the turning points of the frequency in the complex time plane 
determines the Bogoliubov coefficients \cite{Pokrovskii:1961}. 
More specifically, given the convergence of contour integrals, 
we need to pick up the turning points in the complex lower-half time plane. 

In the case (\ref{case1}), $X_\mathbf{k}$ are given by 
\begin{align}
X_\mathbf{k}=eE\cdot\frac{m\bar{k}_3}{2\omega_{\mathbf{k},+}^2\bar{k}},
\ eE\cdot C_\mathbf{k},\ eE\cdot D_\mathbf{k}. 
\end{align} 
The explicit from of the validity condition (\ref{semiclassical}) is given by 
\begin{align}
k_\perp\partial_0\theta\gg eE, \quad
m^2\gg eE. 
\label{VC1}\end{align}
The first condition comes from the conditions on the block-off-diagonal elements, 
and the second condition comes from the conditions on the block-diagonal elements. 
See Appendix \ref{A-Validity} for the detail of the derivation. 
We already used the first condition in deriving (\ref{AAPP1}). 
If this condition is not satisfied, the block-off-diagonal elements are not negligible. 

Let us see the turning points in the complex lower-half time plane. 
In the case (\ref{case1}), the lower frequency has two turning points 
\begin{align}
\bar{k}_3(\tau_0^{(1)})&=-\sqrt{(\partial_0\theta+im)^2-k_\perp^2}, \notag\\
\bar{k}_3(\tau_0^{(2)})&=\sqrt{(\partial_0\theta-im)^2-k_\perp^2}, 
\label{TP1}\end{align}
i.e., $\omega_{\mathbf{k},+}(\tau_0^{(i)})=0$, $\text{Im}\ \tau_0^{(i)}<0$. 
Near the turning points, the following identity holds true, 
\begin{align}
eE\cdot\frac{m\bar{k}_3}{2\omega_{\mathbf{k},+}^2\bar{k}}
\simeq \mp i\frac{\partial_0\omega_{\mathbf{k},+}}{2\omega_{\mathbf{k},+}}, 
\end{align}
where the upper and the lower choices of signs correspond to 
the behaviors around $\tau_0^{(1)}$ and $\tau_0^{(2)}$, respectively. 
The equation (\ref{AAPP1}) is thus written as follows 
\begin{align}
\partial_0\begin{pmatrix} \alpha_{\mathbf{k},+} \\ \beta_{\mathbf{k},+} \end{pmatrix} 
\simeq 
\begin{pmatrix} 0 & \mp i\frac{\partial_0\omega_{\mathbf{k},+}}{\omega_{\mathbf{k},+}}e^{2i\Theta_{\mathbf{k},+}} \\
\pm i\frac{\partial_0\omega_{\mathbf{k},+}}{\omega_{\mathbf{k},+}}e^{-2i\Theta_{\mathbf{k},+}} & 0 \end{pmatrix} 
\begin{pmatrix} \alpha_{\mathbf{k},+} \\ \beta_{\mathbf{k},+} \end{pmatrix}. 
\label{AAPP2}\end{align}
We can approximately solve this equation 
by using the behavior of the frequency near the turning points: 
$\omega_{\mathbf{k},+}\propto (\tau-\tau_0^{(i)})^\frac{1}{2}$. 
The solution is shown below in the form of the distribution function 
of produced fermions; i.e., $n_{\mathbf{k},+}\simeq|\beta_{\mathbf{k},+}|^2$. 
See \cite{Berry:1982, Dumlu:2011rr} for the detail of the calculation process. 

For $\tau>\text{Re}\ \tau_0^{(2)}$, the distribution function is given by 
\begin{align}
n_{\mathbf{k},+}(\tau)
\simeq\big|e^{-2i\Theta_{\mathbf{k},+}(\tau_0^{(1)})} 
-e^{-2i\Theta_{\mathbf{k},+}(\tau_0^{(2)})}\big|^2, 
\label{second1}\end{align}
which consists of the contributions 
from the first and the second turning points $\tau_0^{(1)}$, $\tau_0^{(2)}$. 
The minus sign between the first and the second terms comes from the sign difference 
between the upper and the lower choices in (\ref{AAPP2}). 
For $\text{Re}\ \tau_0^{(2)}>\tau>\text{Re}\ \tau_0^{(1)}$, the distribution function is given by 
\begin{align}
n_{\mathbf{k},+}(\tau)
\simeq\big|e^{-2i\Theta_{\mathbf{k},+}(\tau_0^{(1)})}\big|^2, 
\label{first1}\end{align}
which consists only of the contribution from the first turning point $\tau_0^{(1)}$. 
For $\tau<\text{Re}\ \tau_0^{(1)}$, the pair production has not been occurred. 
We set the initial time as $\tau_\text{ini}<\text{Re}\ \tau_0^{(1)}$. 

It is convenient to express $\Theta_{\mathbf{k},+}(\tau_0^{(i)})$ as the integral over $\bar{k}_3$, 
\begin{align}
\Theta_{\mathbf{k},+}(\tau_0^{(i)})
=\frac{1}{eE}\int^{\bar{k}_3(\tau_0^{(i)})}d\bar{k}_3 
\sqrt{(\sqrt{\bar{k}_3^2+k_\perp^2}-\partial_0\theta)^2+m^2}, 
\label{Theta1}\end{align}
where any real value can be substituted for the lower bound of the integral. 
The choice of real values does not change the distribution function. 
The constant electric field was moved outside the integral. 
Using the imaginary and the real parts of $\Theta_{\mathbf{k},+}(\tau_0^{(i)})$, 
(\ref{second1}) is expressed as follows 
\begin{align}
n_{\mathbf{k},+}(\tau)
&=2\exp(-\mathcal{W}_\mathbf{k})-2\exp(-\mathcal{W}_\mathbf{k})\cos(2\mathcal{P}_\mathbf{k}) \notag\\
&=4\exp(-\mathcal{W}_\mathbf{k}) \sin^2(\mathcal{P}_\mathbf{k}), 
\label{second2}\end{align}
and (\ref{first1}) is expressed as follows 
\begin{align}
n_{\mathbf{k},+}(\tau)
=\exp(-\mathcal{W}_\mathbf{k}), 
\label{first2}\end{align}
where the weight and the phase factors are given by 
\begin{align}
\mathcal{W}_\mathbf{k}
=-4\text{Im}\ \Theta_{\mathbf{k},+}(\tau_0^{(1)}) 
=-4\text{Im}\ \Theta_{\mathbf{k},+}(\tau_0^{(2)}), 
\label{weight1}\end{align}
\begin{align}
\mathcal{P}_\mathbf{k}=\text{Re}\ \Theta_{\mathbf{k},+}(\tau_0^{(1)})
-\text{Re}\ \Theta_{\mathbf{k},+}(\tau_0^{(2)}). 
\label{phase1}\end{align}
As pointed out in \cite{Dumlu:2011rr}, 
if there exist more than one turning points in the complex lower-half time plane, 
they may give an interference effect on the distribution function. 
In (\ref{second1}), the cross term between the contributions from $\tau_0^{(1)}$ and $\tau_0^{(2)}$ 
gives the oscillatory function as 
$-[e^{-2i\Theta_{\mathbf{k},+}(\tau_0^{(1)})}e^{2i\Theta_{\mathbf{k},+}^*(\tau_0^{(2)})}+\text{(h.c.)}]
=-2\exp(-\mathcal{W}_\mathbf{k})\cos(2\mathcal{P}_\mathbf{k})$. 
This is the second term in the first line of (\ref{second2}). 
More specifically, more than one turning points should have been passed. 
This is why (\ref{first1}), and thus (\ref{first2}), have no interference effect. 

The same formula (\ref{second2}) was proposed as an empirical one in \cite{Domcke:2021fee}. 
We gave the derivation of this analytical formula. 
In contrast to the previous study, the adiabatic mode functions were constructed
corresponding to the $\partial_0\theta$-modified frequencies (\ref{DR}). 
Using the appropriate adiabatic basis, 
we derived the equation for the Bogoliubov coefficients (\ref{Bogoliubov4}). 
This equation clearly shows which frequency modes are related to each matrix element. 
Therefore, we could derive the reduced equation for the lower-frequency modes (\ref{AAPP1}). 
From the reduced equation, 
we derived (\ref{second2}) by picking up the contributions from the turning points. 

We here express $\mathcal{W}_\mathbf{k}$ and $\mathcal{P}_\mathbf{k}$ more explicitly. 
The complex integrals (\ref{Theta1}) are not exactly solvable except in the $k_\perp=0$ limit. 
Instead, we evaluate them up to $\mathcal{O}(k_\perp^2)$, 
and thus obtain the following expressions of $\mathcal{W}_\mathbf{k}$ and $\mathcal{P}_\mathbf{k}$, 
\begin{align}
\mathcal{W}_\mathbf{k}
\simeq\frac{\pi m^2}{eE}\Big[
1+\frac{k_\perp^2}{S(S+\partial_0\theta)}\Big], 
\label{weight2}\end{align}
\begin{align}
\mathcal{P}_\mathbf{k}
\simeq\frac{1}{eE}\Big\{&S\partial_0\theta +m^2\log\frac{S+\partial_0\theta}{m} \notag\\
&+\frac{k_\perp^2\partial_0\theta}{S}
\big[\log\frac{k_\perp(S+\partial_0\theta)}{4S^2}-\frac{1}{2}\big] 
+\frac{k_\perp^2m^2}{S(S+\partial_0\theta)}\log\frac{S+\partial_0\theta}{m}\Big\}, 
\label{phase2}\end{align}
where $S\equiv \sqrt{(\partial_0\theta)^2+m^2}$. 

As $\partial_0\theta$ becomes larger than $m$ and $k_\perp$, 
$\mathcal{W}_\mathbf{k}$ and $\mathcal{P}_\mathbf{k}$ approach the following values 
\begin{align}
\mathcal{W}_\mathbf{k}\to \frac{\pi m^2}{eE}, \quad
\mathcal{P}_\mathbf{k}\to \frac{(\partial_0\theta)^2}{eE}. 
\end{align}
The large axion velocity eliminates the $k_\perp$ dependence from the weight factor. 
In this regard, the presence of the axion velocity enhances the amplitude of the distribution function. 
We emphasize that it also enhances the amplitudes of integral quantities like the induced current. 
From (III.20), we can estimate that the momentum integral over $k_\perp$ induces 
$\sqrt{eE}\cdot \partial_0\theta/m$ (rather than $\sqrt{eE}$) per each power of $k_\perp$. 
We also explain the overall factor of the exponential.
This factor is given by the square of the number of relevant turning points; 
i.e., $4$ in (\ref{second2}) while $1$ in (\ref{first2}).\footnote{
In the absence of the axion velocity, 
the distribution function is given by $n_\mathbf{k}=2\exp[-\pi(m^2+k_\perp^2)/(eE)]$ 
where the overall factor $2$ comes from the spin sum. 
Note that (\ref{second2}) and (\ref{first2}) do not have the spin-sum factor  
because the contribution from the higher-frequency modes is negligible.} 
Furthermore, the distribution function shows the oscillatory behavior 
with the change of the axion velocity. 
In particular, at $\partial_0\theta\simeq \sqrt{n\pi eE}$ ($n\in \mathbf{N}$), 
the distribution function can take a tiny value beyond the enhancement effect on the amplitude. 
In evaluating integral quantities, we may move the sine squared outside the $k_\perp$ integral 
because $\mathcal{P}_\mathbf{k} \simeq (\partial_0\theta)^2/(eE)$ 
for $k_\perp\lesssim \sqrt{eE}\cdot\partial_0\theta/m$. 
Although these features were already found in \cite{Domcke:2021fee}, 
we explicitly evaluated the coefficients of the $k_\perp^2$ terms 
of $\mathcal{W}_\mathbf{k}$ and $\mathcal{P}_\mathbf{k}$. 

We here mention the $\partial_0\theta=0$ limit where the standard Schwinger effect should be reproduced. 
The distribution function for the standard Schwinger effect cannot be reproduced 
naively by taking the $\partial_0\theta=0$ limit of (\ref{second2}) with (\ref{weight2}) and (\ref{phase2}). 
In deriving this result, we used (\ref{sizeable}) and (\ref{VC1}),  
which do not allow the $\partial_0\theta=0$ limit. 
Let us get back to the equation for the Bogoliubov coefficients (\ref{Bogoliubov4}). 
At first glance, the $\partial_0\theta=0$ limit of (\ref{Bogoliubov4}) looks different from 
the equation for the Bogoliubov coefficients for the standard Schwinger effect. 
We should recall that the degeneracy of the frequencies recovers 
in the absence of the axion velocity (see (\ref{DR})). 
Making use of the rotational transformation which mixes $u_{\mathbf{k},+}$ and $u_{\mathbf{k},-}$ 
(and also mixes $v_{\mathbf{k},+}$ and $v_{\mathbf{k},-}$), 
we can reproduce the equation for the Bogoliubov coefficients for the standard Schwinger effect. 

\begin{figure}[tbp]
\centering
\includegraphics[width=10cm]{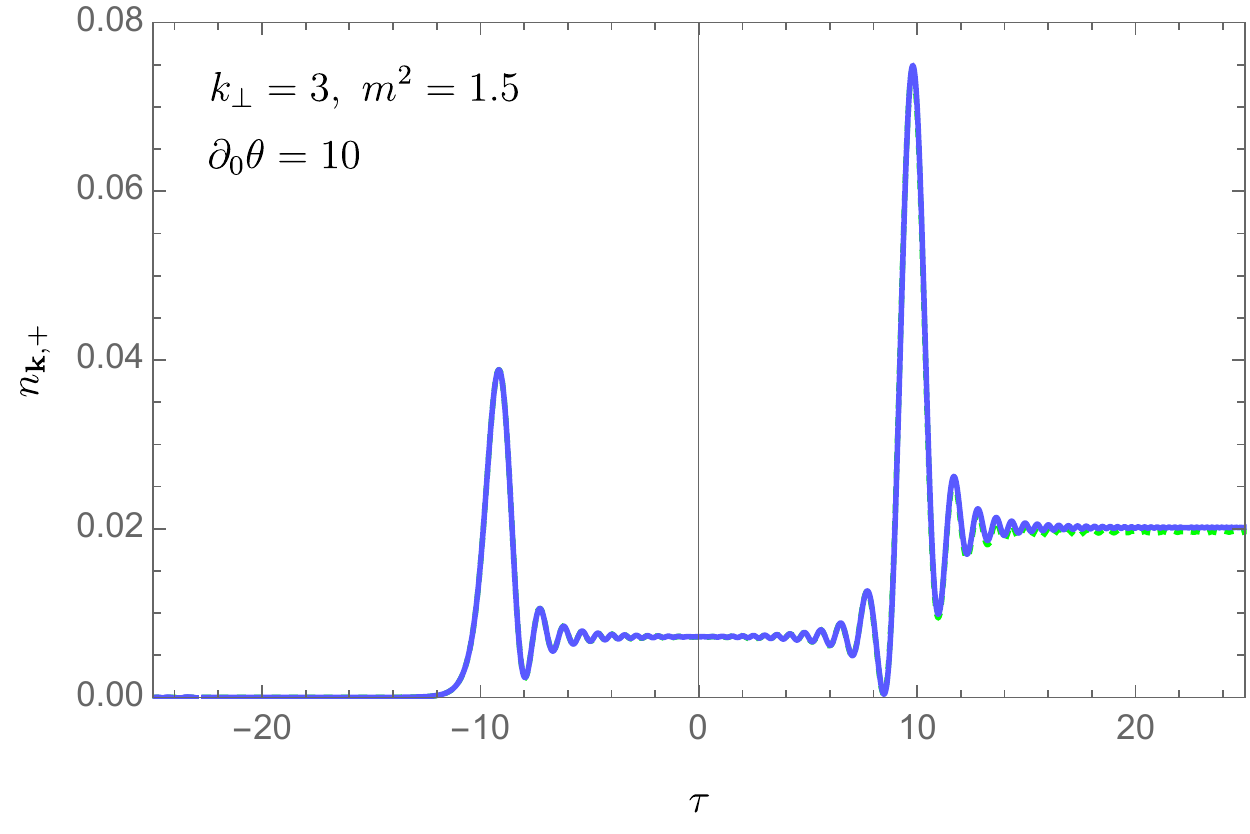}
\caption{$n_{\mathbf{k},+}$ is plotted fixing $\partial_0\theta$, and changing $\tau$. 
The blue solid line is the numerical solution of (\ref{Bogoliubov4}) under $\partial_0^2\theta=0$, 
and the green dotted line is the numerical solution of (\ref{AAPP1}). 
The origin of the time coordinate is set at $\bar{k}_3=0$. 
The parameters are expressed in the $eE=1$ unit.}
\label{Fig1}\end{figure}

In Fig. \ref{Fig1}, we plot the numerical solution of (\ref{Bogoliubov4}) under $\partial_0^2\theta=0$ 
and that of (\ref{AAPP1}), fixing $\partial_0\theta$, and changing $\tau$. 
For the full equation, the distribution function for the lower-frequency modes is given by 
\begin{align}
n_{\mathbf{k},+}=\sum_{\sigma=\pm}
\frac{|\beta_{\mathbf{k},+\sigma}|^2-|\alpha_{\mathbf{k},+\sigma}|^2+\delta_{+\sigma}}{2}, 
\label{full-n}\end{align}
which reduces to (\ref{reduced-n}) in the limit (\ref{reduction}). 
The numerical solution of the full equation is so well described by that of the reduced equation 
that we can barely distinguish between both lines. 
This figure shows that the pair production occurs on two peaks, 
and there exist two steps: 
the first step appears between the two peaks, 
and the second step appears after the later peak. 
It is reasonable to expect that (\ref{first2}) describes the value of the first step, 
and (\ref{second2}) describes the value of the second step. 

\begin{figure}[tbp]
\begin{minipage}{\hsize}
\centering
\includegraphics[width=10cm]{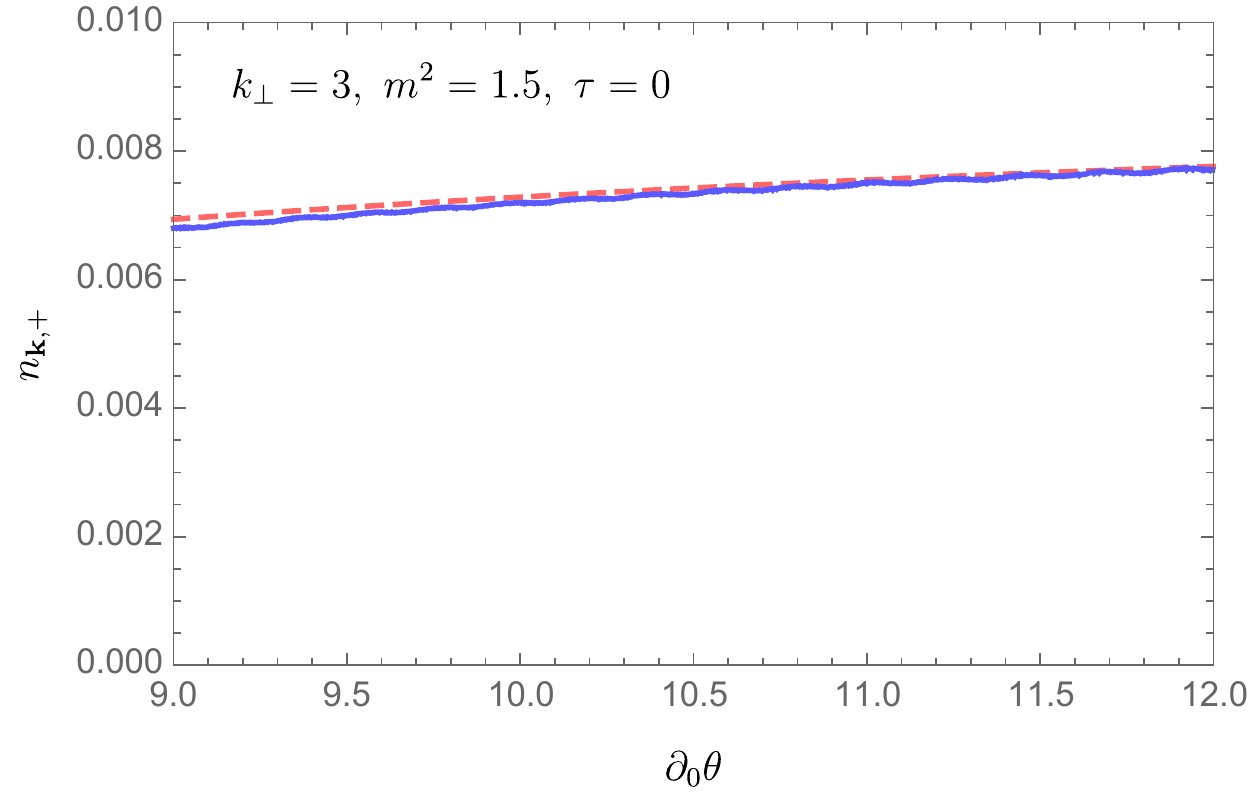}
\end{minipage}\\
\vspace{1em}
\begin{minipage}{\hsize}
\centering
\includegraphics[width=10cm]{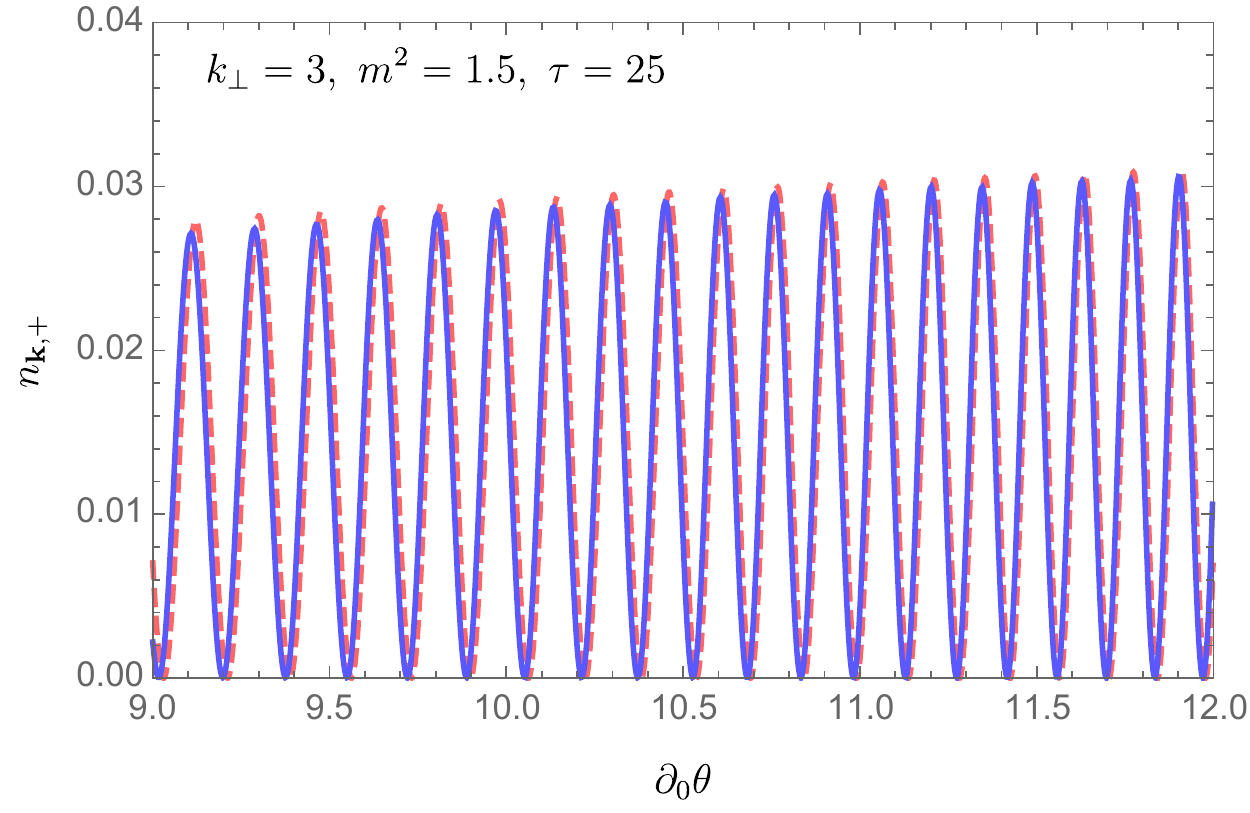}
\end{minipage}
\caption{$n_{\mathbf{k},+}$ is plotted fixing $\tau$, and changing $\partial_0\theta$. 
The blue solid line is the numerical solution of (\ref{Bogoliubov4}) under $\partial_0^2\theta=0$, 
and the red dashed line is the analytical solution. 
In the upper panel, the numerical solution at $\tau=0$ is shown, 
and the analytical solution is given by (\ref{first2}) with (\ref{weight2}). 
In the lower panel, the numerical solution at $\tau=25$ is shown, 
and the analytical solution is given by (\ref{second2}) with (\ref{weight2}) and (\ref{phase2}). 
The parameters are expressed in the $eE=1$ unit.
We here note that the distribution function for the standard Schwinger effect 
$n_\mathbf{k}=2\exp[-\pi(m^2+k_\perp^2)/(eE)]$ takes the tiny value $9.4\times 10^{-15}$.}
\label{Fig2}
\end{figure}

In Fig. \ref{Fig2}, we verify the above expectation: 
we compare the $\partial_0\theta$ dependence of the numerical solution of the full equation 
with that of the analytical solution, at two fixed times. 
In the upper panel, the time is fixed at the value inside the first step ($\tau=0$) for the numerical solution, 
and (\ref{first2}) with (\ref{weight2}) is used as the analytical solution. 
In the lower panel, the time is fixed at the value inside the second step ($\tau=25$) for the numerical solution, 
and (\ref{second2}) with (\ref{weight2}) and (\ref{phase2}) is used as the analytical solution. 
We can confirm that the analytical solution reproduces 
the following features of the numerical solution. 
In the upper panel, $n_{\mathbf{k},+}$ is enhanced with the increase of $\partial_0\theta$. 
In the lower panel, the amplitude of $n_{\mathbf{k},+}$ is enhanced 
with the increase of $\partial_0\theta$, 
and it shows the oscillatory behavior with the change of $\partial_0\theta$. 

\section{Axion-driven pair production}\label{S-ADPP}

We consider the case where the electric field is zero in the whole time range 
while the axion acceleration is nonzero, 
\begin{align}
E=0, \quad \partial_0^2\theta\not=0. 
\label{case2}\end{align}
We eliminate the constant gauge field by shifting the momentum: $\bar{k}_i \to k_i$. 
The axion acceleration is set to be constant and positive for simplicity. 
Furthermore, we restrict the time range such that the axion velocity is kept to be positive.\footnote{
If the axion velocity changes its sign in the time range of our interest, 
we cannot merely call one of $\omega_{\mathbf{k},\pm}$ the lower frequency 
and the other the higher frequency. 
The magnitude relation between them changes with time.} 
Physically speaking, we consider the local time range where the axion velocity changes linearly with time. 

In this case, the pair production is driven by the second term in the right-hand side of (\ref{Bogoliubov4}). 
Since the second term is block-diagonal, we can separately treat 
the contribution from the lower-frequency modes and that from the higher-frequency modes: 
\begin{align}
\alpha_{\mathbf{k},s\sigma}=\alpha_{\mathbf{k},s}\delta_{s\sigma}, \quad
\beta_{\mathbf{k},s\sigma}=\beta_{\mathbf{k},s}\delta_{s\sigma}. 
\end{align}
The unitary condition holds true for each mode, 
\begin{align}
|\alpha_{\mathbf{k},s}|^2+|\beta_{\mathbf{k},s}|^2=1, 
\end{align}
and the distribution function for each mode is given by 
\begin{align}
n_{\mathbf{k},s}=|\beta_{\mathbf{k},s}|^2. 
\end{align}
In the time range where the axion velocity is positive, 
the higher-frequency modes do not cause the pair production. 
We thus consider only the contribution from the lower-frequency modes. 
The equation for $(\alpha_{\mathbf{k},+},\beta_{\mathbf{k},+})$ is given by 
\begin{align}
\partial_0\begin{pmatrix} \alpha_{\mathbf{k},+} \\ \beta_{\mathbf{k},+} \end{pmatrix} 
= 
\partial_0^2\theta\begin{pmatrix} 0 & -\frac{m}{2\omega_{\mathbf{k},+}^2}e^{2i\Theta_{\mathbf{k},+}} \\
\frac{m}{2\omega_{\mathbf{k},+}^2}e^{-2i\Theta_{\mathbf{k},+}} & 0 \end{pmatrix} 
\begin{pmatrix} \alpha_{\mathbf{k},+} \\ \beta_{\mathbf{k},+} \end{pmatrix}. 
\label{ADPP1}\end{align}

We again consider the situation (\ref{semiclassical}) where the semiclassical picture is valid 
on the real time axis. 
In the case (\ref{case2}), $X_\mathbf{k}$ is given by 
\begin{align}
X_\mathbf{k}=\partial_0^2\theta\cdot\frac{m}{2\omega_{\mathbf{k},+}^2}. 
\end{align}
The explicit form of the validity condition is given by 
\begin{align}
m^2\gg \partial_0^2\theta. 
\label{VC2}\end{align}
See Appendix \ref{A-Validity} for the derivation. 
We thus focus on the turning point of the frequency in the complex lower-half time plane. 

In the case (\ref{case2}), the lower-frequency has one turning point 
\begin{align}
\partial_0\theta(\tau_0)=k-im, 
\end{align}
i.e., $\omega_{\mathbf{k},+}(\tau_0)=0$, $\text{Im}\ \tau_0<0$. 
Near the turning point, the following identity holds true, 
\begin{align}
\partial_0^2\theta\cdot\frac{m}{2\omega_{\mathbf{k},+}^2}
\simeq i\frac{\partial_0\omega_{\mathbf{k},+}}{2\omega_{\mathbf{k},+}}. 
\end{align}
The equation (\ref{ADPP1}) is thus written as follows 
\begin{align}
\partial_0\begin{pmatrix} \alpha_{\mathbf{k},+} \\ \beta_{\mathbf{k},+} \end{pmatrix} 
\simeq 
\begin{pmatrix} 0 & -i\frac{\partial_0\omega_{\mathbf{k},+}}{2\omega_{\mathbf{k},+}}e^{2i\Theta_{\mathbf{k},+}} \\
i\frac{\partial_0\omega_{\mathbf{k},+}}{2\omega_{\mathbf{k},+}}e^{-2i\Theta_{\mathbf{k},+}} & 0 \end{pmatrix} 
\begin{pmatrix} \alpha_{\mathbf{k},+} \\ \beta_{\mathbf{k},+} \end{pmatrix}. 
\label{ADPP2}\end{align}
We can approximately solve this equation 
by using the behavior of the frequency near the turning points: 
$\omega_{\mathbf{k},+}\propto (\tau-\tau_0)^\frac{1}{2}$. 

For $\tau>\text{Re}\ \tau_0$ (i.e., $\partial_0\theta>k$), the distribution function is given by 
\begin{align}
n_{\mathbf{k},+}(\tau)\simeq \exp\big[4\text{Im}\ \Theta_{\mathbf{k},+}(\tau_0)\big]. 
\label{single1}\end{align}
For $\tau<\text{Re}\ \tau_0$, the pair production has not been occurred. 
We set the initial time as $\tau_\text{ini}<\text{Re}\ \tau_0$. 
We here explain why the higher-frequency modes do not cause the pair production. 
We can investigate the contribution from the higher-frequency modes 
just by flipping the sign of the axion velocity: $\partial_0\theta\to -\partial_0\theta$. 
The occurrence condition of the pair production is thus given by $-\partial_0\theta>k$. 
Obviously, this condition cannot be satisfied as long as $\partial_0\theta>0$. 

It is convenient to express $\Theta_{\mathbf{k},+}(\tau_0)$ as the integral over $\partial_0\theta$, 
\begin{align}
\Theta_{\mathbf{k},+}(\tau_0)
=\frac{1}{\partial_0^2\theta}\int^{k-im}d(\partial_0\theta) 
\sqrt{(\partial_0\theta-k)^2+m^2}, 
\label{Theta2}\end{align}
where any real value can be substituted for the lower bound of the integral. 
The choice of real values does not change the distribution function. 
The constant axion acceleration was moved outside the integral. 
The complex integral is exactly solvable, 
and thus (\ref{single1}) is given by 
\begin{align}
n_{\mathbf{k},+}(\tau)=\exp\big(-\frac{\pi m^2}{\partial_0^2\theta}\big). 
\label{single2}\end{align}
The weight factor has no momentum dependence. 
We can understand the momentum independence in the following way. 

Let us see that the axion-driven pair production is analogous with the Schwinger effect. 
The dispersion relation for the axion-driven pair production is given by 
\begin{align}
\omega_{\mathbf{k},+}=\sqrt{(k-\partial_0\theta)^2+m^2}, 
\label{analogy1}\end{align}
and that for the Schwinger effect is given by 
\begin{align}
\omega_\mathbf{k}=\sqrt{(k_3-eA)^2+m^2+k_\perp^2}. 
\label{analogy2}\end{align}
Each quantity in (\ref{analogy1}) corresponds to a certain quantity in (\ref{analogy2}): 
$k\leftrightarrow k_3$, $\partial_0\theta\leftrightarrow eA$ 
(i.e., $\partial_0^2\theta\leftrightarrow -eE$), and $m^2\leftrightarrow m^2+k_\perp^2$. 
The first terms in the square roots determine the occurrence condition of the pair production; 
i.e., the pair production occurs when the first terms become zero, 
and the remaining terms appear as the numerators of the weight factors. 
In (\ref{analogy1}), all the momentum dependence is assigned to the first term, 
and thus it does not appear in the distribution function.\footnote{
We here discuss the explicit momentum dependence. 
As seen in (\ref{single3}) and (\ref{single4}), if $\partial_0^2\theta$ and $m$ are changing slowly, 
the distribution function depends on the momentum implicitly via their time evolutions.} 

Before closing this section, we emphasize that 
the semiclassical approach does not use the exact solution of the Dirac equation, 
and thus can be applied to the situation where $\partial_0^2\theta$ and $m$ are not exactly constant 
(recall that $m$ is not constant in an expanding universe). 
We have only to assume that the pair production occurs at the short time scale 
where their time evolutions are negligible. 

Specifically, the turning point is determined as follows 
\begin{align}
\partial_0\theta(\tau)\simeq \partial_0\theta(\text{Re}\ \tau_0)
+(\tau-\text{Re}\ \tau_0)\partial_0^2\theta(\text{Re}\ \tau_0), 
\end{align}
\begin{align}
\tau_0\simeq \text{Re}\ \tau_0-i\frac{m}{\partial_0^2\theta(\text{Re}\ \tau_0)}, 
\end{align}
where $\partial_0\theta(\text{Re}\ \tau_0)=k$. 
The general form of the distribution function is thus given by 
\begin{align}
n_{\mathbf{k},+}(\tau)=\exp\big[-\frac{\pi m^2(\text{Re}\ \tau_0)}
{\partial_0^2\theta(\text{Re}\ \tau_0)}\big]. 
\label{single3}\end{align}
In the physical scale and the cosmic time $t$, it is expressed as follows 
\begin{align}
n_{\mathbf{k},+}(t)=\exp\big[-\frac{\pi m_\text{phys}^2}
{H_\text{phys}(t_\mathbf{k})\dot{\theta}(t_\mathbf{k})+\ddot{\theta}(t_\mathbf{k})}\big], 
\label{single4}\end{align}
where $t_\mathbf{k}= \int^{\text{Re}\ \tau_0}d\tau a$, $H_\text{phys}=\dot{a}/a$, 
and $\dot{}$ means the derivative with respect to the cosmic time. 
This distribution function is consistent with that found in \cite{Adshead:2015kza, Hook:2019zxa}. 

\section{Conclusion and Discussion}\label{S-Con}

We studied the pair production of fermions in a time dependent axion background 
with and without an electric background. 
For this study, we developed the semiclassical approach 
such that the adiabatic mode functions incorporate 
the gauge field and the axion velocity dependence of the dispersion relation. 

The equation (\ref{Bogoliubov4}) summarizes our semiclassical approach, and shows that 
the presence of the axion velocity gives enhancement and interference effects 
on the pair production driven by the electric field (\textit{axion-assisted pair production}), 
and the presence of the axion acceleration itself causes the pair production 
(\textit{axion-driven pair production}). 

For the axion-assisted pair production, we derived the analytical formula (\ref{second2}) and (\ref{first2}) 
which can describe the distribution function of produced fermions 
in the parameter region (\ref{VC1}) for a sizeable axion velocity (\ref{sizeable}). 
This result justifies and improves the empirical formula proposed in \cite{Domcke:2021fee}. 

Also for the axion-driven pair production, we derived the analytical formula (\ref{single2}) 
which can describe the distribution function of produced fermions in the parameter region (\ref{VC2}). 
The consistent results were found in the context of the axion inflation \cite{Adshead:2015kza}, 
and the cosmological collider \cite{Hook:2019zxa}. 

There are some applications of our results to phenomenologically interesting cosmological scenarios. 
The axion-assisted pair production can be important in the scenario of the magnetogenesis 
by the axion inflation \cite{Turner:1987bw, Garretson:1992vt, Anber:2006xt, Barnaby:2011qe, 
Domcke:2018eki, Kamada:2019uxp, Domcke:2019qmm}. 
In the original paper of the axion-assisted pair production \cite{Domcke:2021fee}, 
the authors discussed that the axion-assisted pair production results in a further reduction 
of the gauge field production during the axion inflation. 
The axion-driven pair production can be important, 
e.g., in the kinetic misalignment mechanism, which was recently proposed by \cite{Co:2019wyp, Co:2019jts}. 
This mechanism provides large axion velocity and axion acceleration in the early Universe, 
which may result in significant axion-driven pair production. 

Recalling that the Schwinger effect puts an upper bound on the magnitude of the static electric field, 
we conclude that the axion-driven pair production puts an upper bound 
on the magnitude of the axion acceleration. 
If an axion couples to a light fermion, it cannot be accelerated faster than of order the fermion mass squared.  This may strongly restrict model constructions in some cosmological scenarios, 
including the kinetic misalignment mechanism 
\cite{Co:2019wyp, Co:2019jts, Co:2020dya, Co:2020xlh, Co:2020jtv, Co:2021rhi} 
and the axion inflation \cite{Freese:1990rb, Adshead:2015kza, Nomura:2017ehb}. 

For quantitative study of the influence of the pair production, 
we need to evaluate physical quantities like the induced current and the energy-momentum tensor, 
which include the Bogoliubov coefficients as their integrands. 
We expect that the usage of our adiabatic basis enables the simple expression of these quantities; 
i.e., they are diagonalized with respect to the lower and the higher frequency modes. 
This expectation has been confirmed at least for the induced current, 
detailed calculations of which will be published elsewhere. 

\begin{acknowledgments}
We would like to thank V. Domcke, Y. Ema, and K. Mukaida 
for reading the manuscript and for valuable comments.
H. K. would like to thank the organizers of the YITP workshop on Strings and Fields 2021. 
The discussion during the workshop improved the quality of this work. 
H. K. and M. Y. were supported by Leading Initiative for Excellent Young Researchers, MEXT, Japan. 
M. Y. was also supported by JSPS KAKENHI Grant Numbers 20H0585, 20K22344, and 21K13910. 
\end{acknowledgments}

\appendix

\section{Notation of gamma matrices}\label{A-Notation}

The gamma matrices are defined by the anti-commutation relation 
\begin{align}
\{\gamma^\mu,\gamma^\nu\}=-2\eta^{\mu\nu}I, 
\end{align}
where $I$ is the identity matrix. 
In this paper, we use the chiral representation 
\begin{align}
\gamma^0=\begin{pmatrix} O & I \\ I & O \end{pmatrix}, \quad
\gamma^i=\begin{pmatrix} O & \sigma^i \\ -\sigma^i & O \end{pmatrix}, 
\end{align}
where $O$ is the zero matrix and $\sigma^i$ are the Pauli matrices 
\begin{align}
\sigma^1=\begin{pmatrix} 0 & 1 \\ 1 & 0 \end{pmatrix}, \quad
\sigma^2=\begin{pmatrix} 0 & -i \\ i & 0 \end{pmatrix}, \quad
\sigma^3=\begin{pmatrix} 1 & 0 \\ 0 & -1 \end{pmatrix}. 
\end{align}
The fifth gamma matrix is defined as follows 
\begin{align}
\gamma^5\equiv i\gamma^0\gamma^1\gamma^2\gamma^3
=\begin{pmatrix} -I & O \\ O & I \end{pmatrix}, 
\end{align}
and it anti-commutes with all the gamma matrices 
\begin{align}
\gamma^\mu\gamma^5=-\gamma^5\gamma^\mu. 
\end{align}

\section{Universality of adiabatic mode functions}\label{A-Universality}

For $\bar{k}\le \partial_0\theta$, 
the frequencies (\ref{DR}) are given by solutions of the following equation, 
\begin{align}
\omega_{\mathbf{k},\pm}^2-\bar{k}^2-m^2 
+(\partial_0\theta)^2-2\partial_0\theta\sqrt{\omega_{\mathbf{k},\pm}^2-m^2}=0. 
\end{align}
Note that unlike for $\bar{k}\ge \partial_0\theta$, 
we consider only the case where the sign of the last term in the left-hand side is minus. 
In this case, both solutions satisfy the positivity of $\sqrt{\omega_{\mathbf{k},\pm}^2-m^2}$. 
If the sign is plus, both solutions do not satisfy the positivity and thus we do not consider that case. 

That is to say, for $\bar{k}\le \partial_0\theta$, 
the positive-frequency mode functions are constructed from the following eigenvectors, 
\begin{align}
2\partial_0\theta(-\gamma^5\omega_{\mathbf{k},\pm}+m\gamma^0\gamma^5)
\hat{\mu}_{\mathbf{k},\pm} 
=-2\partial_0\theta\sqrt{\omega_{\mathbf{k},\pm}^2-m^2}\hat{\mu}_{\mathbf{k},\pm}, 
\label{mu1'}\end{align}
\begin{align}
\hat{\mu}_{\mathbf{k},+}
&=\begin{pmatrix} \big(\omega_{\mathbf{k},+}-\sqrt{\omega_{\mathbf{k},+}^2-m^2}\big)\xi \\ 
-m\xi \end{pmatrix}, \notag\\
\hat{\mu}_{\mathbf{k},-}
&=\begin{pmatrix} -m\xi \\ 
\big(\omega_{\mathbf{k},-}+\sqrt{\omega_{\mathbf{k},-}^2-m^2}\big)\xi \end{pmatrix},  
\label{mu2'}\end{align}
and the negative-frequency mode functions are constructed from the following eigenvectors, 
\begin{align}
2\partial_0\theta(\gamma^5\omega_{\mathbf{k},\pm}+m\gamma^0\gamma^5)
\hat{\nu}_{\mathbf{k},\pm} 
=-2\partial_0\theta\sqrt{\omega_{\mathbf{k},\pm}^2-m^2}\hat{\nu}_{\mathbf{k},\pm}, 
\label{nu1'}\end{align}
\begin{align}
\hat{\nu}_{\mathbf{k},+}
&=\begin{pmatrix} m\xi \\ 
\big(\omega_{\mathbf{k},+}-\sqrt{\omega_{\mathbf{k},+}^2-m^2}\big)\xi \end{pmatrix}, \notag\\
\hat{\nu}_{\mathbf{k},-}
&=\begin{pmatrix} \big(\omega_{\mathbf{k},-}+\sqrt{\omega_{\mathbf{k},-}^2-m^2}\big)\xi \\
m\xi \end{pmatrix}. 
\label{nu2'}\end{align}

Comparing (\ref{mu2}) and (\ref{nu2}) with (\ref{mu2'}) and (\ref{nu2'}), respectively, 
we can find that the following forms of $\hat{\mu}_{\mathbf{k},\pm}$ and $\hat{\nu}_{\mathbf{k},\pm}$ 
hold true regardless of the sign of $(\bar{k}-\partial_0\theta)$, 
\begin{align}
\hat{\mu}_{\mathbf{k},+}
&=\begin{pmatrix} (\omega_{\mathbf{k},+}+\bar{k}-\partial_0\theta)\xi \\ 
-m\xi \end{pmatrix}, \notag\\
\hat{\mu}_{\mathbf{k},-}
&=\begin{pmatrix} -m\xi \\
(\omega_{\mathbf{k},-}+\bar{k}+\partial_0\theta)\xi \end{pmatrix}, 
\label{mu2''}\end{align}
\begin{align}
\hat{\nu}_{\mathbf{k},+}
&=\begin{pmatrix} m\xi \\
(\omega_{\mathbf{k},+}+\bar{k}-\partial_0\theta)\xi \end{pmatrix}, \notag\\
\hat{\nu}_{\mathbf{k},-}
&=\begin{pmatrix} (\omega_{\mathbf{k},-}+\bar{k}+\partial_0\theta)\xi \\
m\xi \end{pmatrix}. 
\label{nu2''}\end{align}
The adiabatic mode functions (\ref{u+}), (\ref{u-}), (\ref{v+}), and (\ref{v-}) 
are constructed from these universal forms, 
and thus hold true regardless of the sign of $(\bar{k}-\partial_0\theta)$. 

\section{Products of adiabatic mode functions and their time derivatives}\label{A-Products}

The explicit form of 
$T_\mathbf{k} \equiv (\partial_0u_{\mathbf{k},+}^\dagger, \partial_0v_{\mathbf{k},+}^\dagger, 
\partial_0u_{\mathbf{k},-}^\dagger, \partial_0v_{\mathbf{k},-}^\dagger) \otimes
(u_{\mathbf{k},+}, v_{\mathbf{k},+}, u_{\mathbf{k},-}, v_{\mathbf{k},-})$ is given by 
\begin{align}
T_\mathbf{k}=
&\begin{pmatrix}
i\omega_{\mathbf{k},+} & 0 & 0 & 0 \\
0 & -i\omega_{\mathbf{k},+} & 0 & 0 \\
0 & 0 & i\omega_{\mathbf{k},-} & 0 \\
0 & 0 & 0 & -i\omega_{\mathbf{k},-} 
\end{pmatrix} \notag\\
&+eE
\begin{pmatrix}
0 &
\frac{m\bar{k}_3}{2\omega_{\mathbf{k},+}^2\bar{k}}e^{2i\Theta_{\mathbf{k},+}} &
C_\mathbf{k}e^{i(\Theta_{\mathbf{k},+}-\Theta_{\mathbf{k},-})} &
-D_\mathbf{k}e^{i(\Theta_{\mathbf{k},+}+\Theta_{\mathbf{k},-})} \\
-\frac{m\bar{k}_3}{2\omega_{\mathbf{k},+}^2\bar{k}}e^{-2i\Theta_{\mathbf{k},+}} &
0 &
-D_\mathbf{k}e^{-i(\Theta_{\mathbf{k},+}+\Theta_{\mathbf{k},-})} &
-C_\mathbf{k}e^{-i(\Theta_{\mathbf{k},+}-\Theta_{\mathbf{k},-})} \\
-C_\mathbf{k}e^{-i(\Theta_{\mathbf{k},+}-\Theta_{\mathbf{k},-})} &
D_\mathbf{k}e^{i(\Theta_{\mathbf{k},+}+\Theta_{\mathbf{k},-})} &
0 &
\frac{m\bar{k}_3}{2\omega_{\mathbf{k},-}^2\bar{k}}e^{2i\Theta_{\mathbf{k},-}} \\
D_\mathbf{k}e^{-i(\Theta_{\mathbf{k},+}+\Theta_{\mathbf{k},-})} &
C_\mathbf{k}e^{i(\Theta_{\mathbf{k},+}-\Theta_{\mathbf{k},-})} &
-\frac{m\bar{k}_3}{2\omega_{\mathbf{k},-}^2\bar{k}}e^{-2i\Theta_{\mathbf{k},-}} &
0  
\end{pmatrix} \notag\\
&+\partial_0^2\theta
\begin{pmatrix}
0 & -\frac{m}{2\omega_{\mathbf{k},+}^2}e^{2i\Theta_{\mathbf{k},+}} & 0 & 0 \\
\frac{m}{2\omega_{\mathbf{k},+}^2}e^{-2i\Theta_{\mathbf{k},+}} & 0 & 0 & 0 \\
0 & 0 & 0 & \frac{m}{2\omega_{\mathbf{k},-}^2}e^{2i\Theta_{\mathbf{k},-}} \\
0 & 0 & -\frac{m}{2\omega_{\mathbf{k},-}^2}e^{-2i\Theta_{\mathbf{k},-}} & 0 
\end{pmatrix}, 
\label{Products}\end{align}
where $C_\mathbf{k}$ and $D_\mathbf{k}$ are given by 
\begin{align}
C_\mathbf{k}
&=\frac{m\big[(\omega_{\mathbf{k},+}+\bar{k}-\partial_0\theta)
+(\omega_{\mathbf{k},-}+\bar{k}+\partial_0\theta)\big]}
{4\sqrt{\omega_{\mathbf{k},+}\omega_{\mathbf{k},-}
(\omega_{\mathbf{k},+}+\bar{k}-\partial_0\theta)(\omega_{\mathbf{k},-}+\bar{k}+\partial_0\theta)}}
\frac{k_\perp}{\bar{k}^2}, 
\label{C}\end{align}
\begin{align}
D_\mathbf{k}
&=\frac{(\omega_{\mathbf{k},+}+\bar{k}-\partial_0\theta)(\omega_{\mathbf{k},-}+\bar{k}+\partial_0\theta)-m^2}
{4\sqrt{\omega_{\mathbf{k},+}\omega_{\mathbf{k},-}
(\omega_{\mathbf{k},+}+\bar{k}-\partial_0\theta)(\omega_{\mathbf{k},-}+\bar{k}+\partial_0\theta)}}
\frac{k_\perp}{\bar{k}^2}. 
\label{D}\end{align}
In the right-hand side of (\ref{Products}), 
the first, the second, and the third terms come from the time derivatives of 
$\Theta_{\mathbf{k},\pm}$, $A_\mu$, and $\partial_0\theta$, respectively. 
The first term is canceled out in the equation for the Bogoliubov coefficients (\ref{Bogoliubov4}). 

\section{Validity condition of semiclassical picture}\label{A-Validity}

In the case (\ref{case1}), $(X_\mathbf{k}/\omega_{\mathbf{k},+})^2$ 
and $|\partial_0X_\mathbf{k}/\omega_{\mathbf{k},+}^2|$ can be large 
when $\bar{k}$ and $\omega_{\mathbf{k},+}$ in the denominators take their minimum values.  
Therefore, the validity condition of the semiclassical picture (\ref{semiclassical}) is satisfied 
as long as the reference quantities are much smaller than unity 
at $\bar{k}_3=0$ and $\bar{k}=\partial_0\theta$. 

At $\bar{k}_3=0$, the reference quantities are given by 
\begin{align}
&\Big(\frac{eE}{\omega_{\mathbf{k},+}}\cdot \frac{m\bar{k}_3}{2\omega_{\mathbf{k},+}^2\bar{k}}\Big)^2
=0, \notag\\
&\Big(\frac{eE}{\omega_{\mathbf{k},+}}\cdot C_\mathbf{k}\Big)^2
\simeq \frac{(eE)^2}{4k_\perp^2(\partial_0\theta)^2}, \quad
\Big(\frac{eE}{\omega_{\mathbf{k},+}}\cdot D_\mathbf{k}\Big)^2
\simeq \frac{(eE)^2\cdot m^2}{4(\partial_0\theta)^6},
\end{align}
\begin{align}
&\Big|\frac{eE}{\omega_{\mathbf{k},+}^2}
\cdot \partial_0\big(\frac{m\bar{k}_3}{2\omega_{\mathbf{k},+}^2\bar{k}}\big)\Big|
\simeq \frac{(eE)^2\cdot m}{2k_\perp(\partial_0\theta)^4}, \notag\\
&\Big|\frac{eE}{\omega_{\mathbf{k},+}^2}\cdot \partial_0C_\mathbf{k}\Big|
=0, \quad
\Big|\frac{eE}{\omega_{\mathbf{k},+}^2}\cdot \partial_0D_\mathbf{k}\Big|
=0. 
\end{align}
Recall that we consider a sizeable axion velocity $\partial_0\theta\gg k_\perp,m$. 
At $\bar{k}=\partial_0\theta$, the reference quantities are given by 
\begin{align}
&\Big(\frac{eE}{\omega_{\mathbf{k},+}}\cdot \frac{m\bar{k}_3}{2\omega_{\mathbf{k},+}^2\bar{k}}\Big)^2
\simeq\frac{(eE)^2}{4m^4}, \notag\\
&\Big(\frac{eE}{\omega_{\mathbf{k},+}}\cdot C_\mathbf{k}\Big)^2
\simeq \frac{(eE)^2\cdot k_\perp^2}{8m^2(\partial_0\theta)^4}, \quad
\Big(\frac{eE}{\omega_{\mathbf{k},+}}\cdot D_\mathbf{k}\Big)^2
\simeq \frac{(eE)^2\cdot k_\perp^2}{8m^2(\partial_0\theta)^4},
\end{align}
\begin{align}
&\Big|\frac{eE}{\omega_{\mathbf{k},+}^2}
\cdot \partial_0\big(\frac{m\bar{k}_3}{2\omega_{\mathbf{k},+}^2\bar{k}}\big)\Big|
\simeq \frac{(eE)^2\cdot k_\perp^2}{2m^3(\partial_0\theta)^3}, \notag\\
&\Big|\frac{eE}{\omega_{\mathbf{k},+}^2}\cdot \partial_0C_\mathbf{k}\Big|
\simeq \frac{(eE)^2\cdot k_\perp}{4\sqrt{2}m^3(\partial_0\theta)^2}, \quad
\Big|\frac{eE}{\omega_{\mathbf{k},+}^2}\cdot \partial_0D_\mathbf{k}\Big|
\simeq\frac{(eE)^2\cdot k_\perp}{4\sqrt{2}m^3(\partial_0\theta)^2}. 
\end{align}
We thus obtain the explicit form of the validity condition, 
\begin{align}
k_\perp\partial_0\theta\gg eE, \quad
m^2\gg eE. 
\end{align}

In the case (\ref{case2}), $(X_\mathbf{k}/\omega_{\mathbf{k},+})^2$ 
and $|\partial_0X_\mathbf{k}/\omega_{\mathbf{k},+}^2|$ can be large 
when $\omega_{\mathbf{k},+}$ in the denominators takes its minimum value.  
Therefore, the reference quantities have to be much smaller than unity at $k=\partial_0\theta$. 

At $k=\partial_0\theta$, the reference quantities are given by 
\begin{align}
\Big(\frac{\partial_0^2\theta}{\omega_{\mathbf{k},+}}\cdot \frac{m}{2\omega_{\mathbf{k},+}^2}\Big)^2
=\frac{(\partial_0^2\theta)^2}{4m^4}, 
\end{align}
\begin{align}
\Big|\frac{\partial_0^2\theta}{\omega_{\mathbf{k},+}^2}
\cdot \partial_0\big(\frac{m}{2\omega_{\mathbf{k},+}^2}\big)\Big|
=0.  
\end{align}
We thus obtain the explicit form of the validity condition, 
\begin{align}
m^2\gg \partial_0^2\theta. 
\end{align}


\begin{thebibliography}{99}

\bibitem{Schwinger:1951nm} 
J. S. Schwinger, 
Phys. Rev. \textbf{82}, 664 (1951).

\bibitem{Turner:1987bw}
M. S. Turner and L. M. Widrow,
Phys. Rev. D \textbf{37}, 2743 (1988).

\bibitem{Garretson:1992vt}
W. D. Garretson, G. B. Field, and S. M. Carroll,
Phys. Rev. D \textbf{46}, 5346 (1992) [arXiv:hep-ph/9209238].

\bibitem{Anber:2006xt}
M. M. Anber and L. Sorbo,
J. Cosmol. Astropart. Phys. \textbf{10} (2006), 018 [arXiv:astro-ph/0606534].

\bibitem{Barnaby:2011qe}
N. Barnaby, E. Pajer, and M. Peloso,
Phys. Rev. D \textbf{85}, 023525 (2012) [arXiv:1110.3327 [astro-ph.CO]].

\bibitem{Domcke:2018eki}
V. Domcke and K. Mukaida,
J. Cosmol. Astropart. Phys. \textbf{11} (2018), 020 [arXiv:1806.08769 [hep-ph]].

\bibitem{Kamada:2019uxp}
K. Kamada and C. S. Shin,
J. High Energy Phys. \textbf{04} (2020), 185 [arXiv:1905.06966 [hep-ph]].

\bibitem{Domcke:2019qmm}
V. Domcke, Y. Ema, and K. Mukaida,
J. High Energy Phys. \textbf{02} (2020), 055 [arXiv:1910.01205 [hep-ph]].

\bibitem{Domcke:2021fee}
V. Domcke, Y. Ema, and K. Mukaida, 
J. High Energy Phys. \textbf{05} (2021) 001 [arXiv:2101.05192 [hep-ph]].

\bibitem{Brezin:1970xf}
E. Brezin and C. Itzykson,
Phys. Rev. D \textbf{2}, 1191 (1970).

\bibitem{Popov:1972}
V. S. Popov,
Zh. Eksp. Teor. Fiz. 62, 1248 (1972) [Sov. Phys. JETP 35, 659 (1972)].

\bibitem{Berry:1982}
M. V. Berry, J. Phys. A \textbf{15}, 3693 (1982).

\bibitem{Kluger:1998bm}
Y. Kluger, E. Mottola, and J. M. Eisenberg,
Phys. Rev. D \textbf{58}, 125015 (1998) [arXiv:hep-ph/9803372 [hep-ph]].

\bibitem{Dumlu:2011rr}
C. K. Dumlu and G. V. Dunne, 
Phys. Rev. D \textbf{83}, 065028 (2011) [arXiv:1102.2899 [hep-th]].

\bibitem{Adshead:2015kza}
P. Adshead and E. I. Sfakianakis,
J. Cosmol. Astropart. Phys. \textbf{11} (2015), 021 [arXiv:1508.00891 [hep-ph]].

\bibitem{Hook:2019zxa}
A. Hook, J. Huang, and D. Racco, 
J. High Energy Phys. \textbf{01} (2020), 105 [arXiv:1907.10624 [hep-ph]].

\bibitem{Co:2019wyp}
R. T. Co and K. Harigaya,
Phys. Rev. Lett. \textbf{124}, 111602 (2020) [arXiv:1910.02080 [hep-ph]].

\bibitem{Co:2019jts}
R. T. Co, L. J. Hall, and K. Harigaya,
Phys. Rev. Lett. \textbf{124}, 251802 (2020) [arXiv:1910.14152 [hep-ph]].

\bibitem{Pokrovskii:1961} 
V. L. Pokrovskii and I. M. Khalatnikov, 
Zh. Eksp. Teor. Fiz. \textbf{40}, 1713 (1961) [Sov. Phys. JETP \textbf{13}, 1207 (1961)].

\bibitem{Co:2020dya}
R. T. Co, L. J. Hall, K. Harigaya, K. A. Olive, and S. Verner,
J. Cosmol. Astropart. Phys. \textbf{08} (2020), 036 [arXiv:2004.00629 [hep-ph]].

\bibitem{Co:2020xlh}
R. T. Co, L. J. Hall, and K. Harigaya,
J. High Energy Phys. \textbf{01} (2021), 172 [arXiv:2006.04809 [hep-ph]].

\bibitem{Co:2020jtv}
R. T. Co, N. Fernandez, A. Ghalsasi, L. J. Hall, and K. Harigaya,
J. High Energy Phys. \textbf{21} (2020), 017 [arXiv:2006.05687 [hep-ph]].

\bibitem{Co:2021rhi}
R. T. Co, K. Harigaya, and A. Pierce,
J. High Energy Phys. \textbf{12} (2021), 099 [arXiv:2104.02077 [hep-ph]].

\bibitem{Freese:1990rb}
K. Freese, J. A. Frieman, and A. V. Olinto,
Phys. Rev. Lett. \textbf{65}, 3233 (1990).

\bibitem{Nomura:2017ehb}
Y. Nomura, T. Watari, and M. Yamazaki,
Phys. Lett. B \textbf{776}, 227 (2018) [arXiv:1706.08522 [hep-ph]].

\end{thebibliography}
\end{document}